\documentclass[10pt,nofootinbib]{revtex4}
\begin{document}

\title{{\Large ``Forget time"}\\
{\normalsize\rm\em Essay written for the FQXi contest on the Nature of Time}}
\author{Carlo Rovelli}
\date{\small August 2008}
\begin{abstract}
\noindent Following a line of research that I have developed for several years, I argue that
the best strategy for understanding quantum gravity is to build a picture of the
physical world where the notion of time plays no role at all.  I summarize
here this point of view, explaining \emph{why} I think that in a fundamental description of 
nature  we must ``forget time", and \emph{how} this can be done in the classical
and in the quantum theory.  The idea is to develop a formalism that treats
dependent and independent variables on the same footing. In short, I
propose to interpret mechanics as a theory of relations between variables, 
rather than  the theory of the evolution of variables in time.
\end{abstract}
\maketitle

\section{The need to forget time}

General relativity has changed our understanding of space and time. The efforts to find a theory capable of describing the expected quantum properties of  gravity forces us to fully confront this change, and perhaps even push it further.  The spacetime of general relativity, indeed, (a 4d pseudo-riemannian space) is likely to be just a classical approximation  that loses its meaning in the quantum theory, for the same reason the {\em trajectory} of a particle does (see for instance \cite{Ashtekar:2004vs}).   How should we then think about time in the future quantum theory of gravity? 

Here I argue for a possible answer to this question.  The answer I defend is that we must forget the notion of time altogether, and build a quantum theory of gravity where this notion does not appear at all. The notion of time familiar to us may then be reconstructed in special physical situations, or within an approximation, as is the case for a number of familiar physical quantities that disappear when moving to a deeper level of description (for instance: the ``surface of a liquid" disappears when going to the atomic level, or ``temperature" is a notion that makes sense only in certain physical situations and when there are enough degrees of freedom.) 

I have argued for this point of view in the past in a number of papers [2-13], and in the book \cite{book}.  However, I have never tried to articulate this point of view in a compact, direct, and self-contained way.  I take therefore the opportunity of the FQXi contest for doing so.  In the following I explain in detail what I mean by ``forgetting time", why I think it is necessary to do so in order to construct a quantum theory of gravity, and how I think it is possible.  The discussion here is general, and focused on the notion of time: I will not enter at all in the technical complications associated with the definition of a quantum theory of gravity (on this, see  \cite{book,loop} for one possible approach), but only discuss the change in the notion of time that I believe is required for a deeper understanding of the world, in the light of what we have learned about Nature with general relativity and quantum theory. 

I am aware that the answer I outline here is only one among many possibilities. Other authors have argued that the notion of time is irreducible, and cannot be left out of a fundamental description of Nature in the way I propose.  Until our theoretical and experimental investigations tell us otherwise, I think that what is important is to put the alternatives clearly on the table, and extensively discuss their rationale and consistency. It is in this spirit that I present here the timeless point of view. 

\section{Time as the independent parameter for the evolution}

Until the Relativity revolution, the notion of time played a clear and uncontroversial role in physics. In pre-relativistic physics, time is one of the fundamental notions in terms of which physics is built.  Time is assumed to ``flow", and mechanics is understood as the science describing the laws of the change of the physical systems in time.  Time is described by a real variable $t$.  If the state of a physical system can be described by a set of physical variables $a_n$, then the evolution of the system is described by the {\em functions} $a_n(t)$.  The laws of motion are differential equations for these quantities.\footnote{The idea that physical laws must express the necessary change of the systems \emph{in time} is very old. It can be traced to the most ancient record we have of the idea of natural law: a fragment from the VI century BC's Greek philosopher Anaximander, that says: {\em All things transform into one another, [...]  following the necessity [...] and accor
 ding to the order of time}. }

A comprehensive theoretical framework for mechanics is provided by the Hamiltonian theory.  In this, the state of a system with $n$ degrees of freedom is described by 2n variables $(q_i,p^i)$, the coordinates of the phase space; the dynamics is governed by the Hamiltonian $H(q_i,p^i)$, which is a function on the phase space; and the equations of motion are given by the Hamilton equations
$dq_i/dt=\partial H/\partial p^i, dp^i/dt=-\partial H/\partial q_i$. It is clear that the physical variable $t$ is an essential ingredient of the picture and plays a very special role in this framework.
This role can be summarized as follows:  in pre-relativistic mechanics, time is a special physical quantity, whose value is measured by physical clocks, that plays the role of the independent variable of physical evolution. 

\section{Mechanics is about relations between variables}
The Relativity revolution has modified the notion of time in a number of ways.  I skip here the changes introduced by Special Relativity (in particular, the relativity of simultaneity), which are much discussed and well understood, and concentrate on those  introduced by general relativity.   

In the formalism of general relativity, we can distinguish different notions of time. In particular, we must distinguish the coordinate time $t$ that appears as the argument of the field variable, for instance in $g_{\mu\nu}(x,t)$, from the proper time $s$ measured along a given world line $\gamma=(\gamma^\mu(\tau))$, defined by $s=\int_\gamma d\tau \sqrt{g_{\mu\nu}(\gamma(\tau))d\gamma^\mu/d\tau\; d\gamma^\nu/d\tau}.$

The coordinate time $t$ plays the same role as evolution parameter of the equations of motion as ordinary non-relativistic time.  The equations of motion of the theory are indeed the Einstein equations, which can be seen as second order evolution equations in $t$.  However, the physical interpretation of $t$ is very different from the interpretation of the variable with the same name in the non-relativistic theory.   While non-relativistic time is the observable quantity measured (or approximated) by physical clocks, in general relativity clocks measure $s$ along their worldline, not $t$. The relativistic coordinate $t$ is a freely chosen label with no direct physical interpretation.  This is a well known consequence of the invariance of the Einstein equations under general changes of coordinates.  The physical content of a solution of Einstein's equations is not in its dependence on $t$, but rather in what remains once the dependence on $t$ (and $x$) has been factored away. 

What is then the physical content of a solution of Einstein's equations, and how is evolution described in this context?  To answer, consider what is actually measured in general relativistic experiments.  Here are some typical examples:
\begin{description}
\item[Two clocks.]  Consider a clock at rest on the surface of the Earth and a clock on a satellite in orbit around the Earth. Call $T_1$ and $T_2$ the readings of the two clocks.  Each measures the proper time along its own worldline, in the Earth gravitational field.  The two readings can be taken repeatedly (say at each passage of the satellite over the location of the Earth clock.)  Let $(T'_1,T'_2), (T''_1,T''_2),..., (T^n_1,T^n_2), ... $ be the sequence of readings.   This can be compared with the theory.   More precisely, given a solution of the Einstein eqs (for gravitational field, Earth and satellite), the theory predicts the value of $T_2$ that will be associated to each value of $T_1$. Or vice versa. 
\item[Solar system.] The distances between the Earth and the other planets of the solar system can be measured with great accuracy today (for instance by the proper time on Earth between the emission of a laser pulse and the reception of its echo from the planet.) Call $d_p$ the distance from the planet $p$, measure these distances repeatedly, and write a table of quantities  $d'_p, d''_p, d'''_p, ... d^n_p, ... $. Then we can ask if these fit a sequence predicted by the theory. Again, given a solution of Einstein's equations, the theory predicts which $n$-tuplets $(d_p)$ are possible. 
\item[Binary pulsar.] A binary pulsar is a system of two stars rotating around each other, where one of the two is a pulsar, namely a star that sends a regular pulsating beep. The frequency at which we receive the beeps is modulated by the Doppler effect: it is higher when the pulsar is in the phase of the revolution where it moves towards us.  Therefore we receive a pulsating signal with  frequency increasing and decreasing periodically.  Let $n$ be the number of pulses we receive and $N$ the number of periods.   We can plot the two against each other and describe the increase of $N$ as a function of $n$ (or vice versa).\footnote{By doing so for 10 years for the binary pulsar PSR 1913+16, R Hulse and J Taylor have been able to measure the decrease of the period of revolution of this system, which is indirect evidence for gravitational waves emission, and have won the 1993 Nobel Prize in Physics.}
\end{description}
The moral of these examples is that in general relativistic observations {\em there is no preferred independent time variable}.   What we measure are a number of variables, all on equal footing, and their \emph{relative} evolution.   The first example is particularly enlightening:  notice that it can be equally read it as the evolution of the variable $T_1$ as a function of the variable $T_2$, or viceversa.  Which of the two is the independent variable here?\footnote{Just in case the reader is tempted to take the clock on Earth as a ``more natural" definition of time, recall that of the two clocks the only one in free fall is the orbiting one.}

The way evolution is treated in general relativity, is therefore more subtle than in pre-relativistic theory.   Change is \emph{not} described as evolution of physical variables as a function of a preferred independent observable time variable.  Instead, it is described in terms of a functional relation among equal footing variables (such as the two clock readings $(T_1,T_2)$ or the various planet distances $d_p$, or the two observed quantities $n$ and $N$ of the binary pulsar system).   

In general relativity, there isn't a \emph{preferred} and \emph{observable} quantity that plays the role of independent parameter of the evolution, as there is in non-relativistic mechanics.   General relativity describes {\em the relative evolution of observable quantities}, not the evolution of quantities as functions of a preferred one.   To put it pictorially: with general relativity we have understood that the Newtonian ``big clock" ticking away the ``true universal time"  is not there. 

I think that this feature of general relativity must be taken seriously in view of the problems raised by the attempts to write a quantum theory in gravity.  In the classical theory, the 4-dimensional spacetime continuum continues to give us a good intuition of a flowing time.   But in the quantum theory the 4-dimensional continuum is most likely not anymore there (although some aspect of this could remain \cite{Baez:1997zt}), and notions such as ``the quantum state of the system at time $t$" are quite unnatural in a general relativistic context.   I think we can get a far more effective grasp on the system if we forget notions such as ``evolution in the observable time $t$" or  ``the state of the system at time $t$",  we take the absence of a preferred independent time notion seriously, and we re-think mechanics as a theory of relative evolution of variables, rather than a theory of the evolution of variables in time.   In section \ref{timeless} below I show how this can be 
 done.  Before that, however, allow me to go back to non-relativistic mechanics for a moment. 

\subsection{Back to Galileo and Newton}

Clockmaking owes a lot to Galileo Galilei, who discovered that the small oscillations of a pendulum are isochronous.  The story goes that Galileo was in the cathedral of Pisa during a religious function, watching a big chandelier oscillating.  Using his own pulse as a clock, Galileo discovered that there was the same number of pulses within each oscillation of the chandelier.   Sometime later, pendulum clocks become widespread, and doctors started using them to check the pulse of the ill.    What is going on here?  The oscillations of a pendulum are measured against pulse, and pulse is measured again the pendulum!  How do we know that a clock measures time, if we can only check it against another clock? 

Isaac Newton provides a nice clarification of this issue in the {\em Principia}. According to Newton, we never directly measure the \emph{true} time variable $t$.  Rather, we always construct  devises, the ``clocks" indeed, that have observable quantities (say, the angle $\beta$ between the clock's hand and the direction of the digit ``12"), that move proportionally to the true time, within an approximation good enough for our purposes.   In other words, we can say, following Newton, that what we can observe are the system's quantities $a_i$ and the clock's quantity $\beta$, and their relative evolution, namely the functions $a_i(\beta)$; but we describe this in our theory by assuming the existence of a ``true" time variable $t$.  We can then write evolution equations $a_i(t)$ and $\beta(t)$, and compare these with the observed change of $a_i$ with the clock's hand $a_i(\beta)$.  

Thus, it is true \emph{also} in non-relativistic mechanics that what we measure is only relative evolution between variables.   But it turns out to be convenient to \emph{assume}, with Newton, that there exist a background variable $t$, such that all observables quantities evolve with respect to it, and equations are simple when written with respect to it.   

What I propose to do in the following is simply to drop this assumption. 

\section{Formal structure of timeless mechanics}\label{timeless}

In its conventional formulation, mechanics describes
the evolution of states and observables in time.  This evolution is
governed by a Hamiltonian.  (This is also true for special relativistic
theories and field theories: evolution is governed by a representation of the Poincar\'e
group, and one of the generators of this group is the Hamiltonian.)  
This conventional formulation is
not sufficiently broad, because general relativistic systems --in
fact, the world in which we live-- do not fit in this conceptual
scheme.  Therefore we need a more general formulation of mechanics
than the conventional one.  This formulation must be based on notions
of ``observable" and ``state" that maintain a clear meaning in a
general relativistic context.  I describe here how such a formulation can be defined.\footnote{The
formalism defined here is based on well known works. 
For instance, Arnold \index{Arnold}
\cite{Arnold} identifies the (presymplectic) space with coordinates
$(t,q^i,p_{i})$ (time, lagrangian variables and their momenta) as the
natural home for mechanics.  Souriau has developed a beautiful and
little known relativistic formalism \index{Souriau} \cite{Souriau}. 
Probably the first to consider the point of view used here was
Lagrange \index{Lagrange} himself, in pointing out that the most
convenient definition of ``phase space" is the space of the physical
motions \index{phase space!relativistic} \cite{Lagrange}.  Many of the
tools used below are also used in hamiltonian treatments of general
covariant theories as constrained systems, on the basis of Dirac's theory, 
but generally within a rather obscure interpretative cloud, which I make an
effort here to simplify and clarify.}

\subsection{The harmonic oscillator revisited}

Say we want to describe the small oscillations of a pendulum.  To this
aim, we need \emph{two} measuring devices.  A clock and a device that
reads the elongation of the pendulum.  Let $\alpha$ be  the reading of the device measuring
the elongation of the pendulum and  $\beta$ be the reading of the
clock (say the angle between the clock's hand and the ``12".) Call the variables
$\alpha$ and $\beta$  the \emph{partial observables}
of the pendulum.   

A physically relevant observation is a reading of $\alpha$ and 
$\beta$, \emph{together}.  Thus, an observation yields a pair
$(\alpha,\beta)$.  Call a pair obtained in this manner an {\em event}. 
Let $\cal C$ be the two-dimensional space with coordinates $\alpha$ and
$\beta$.  Call $\cal C$ the {\em event space} of
the pendulum. 

Experience shows that we can find mathematical laws characterizing {\em
sequences} of events.  (This is the reason we can do science.)  These
laws have the following form.  Call a unparametrized curve $\gamma$ in
$\cal C$ a {\em motion\/} of the system.  Perform a sequence of
measurements of pairs $(\alpha,\beta)$, and find that the points
representing the measured pairs sit on a motion $\gamma$.  Then we say
that $\gamma$ is a {\em physical motion\/}.   We express
a motion as a relation in $\cal C$
\begin{equation}
    f(\alpha,\beta)=0.
\label{uno}
\end{equation}
Thus a motion $\gamma$ is a relation (or a {\em correlation})
between partial observables.

Then, disturb the pendulum (push it with a finger) and repeat the
entire experiment over.  At each repetition of the experiment, a
different motion $\gamma$ is found.  That is, a different mathematical
relation of the form (\ref{uno}) is found.  Experience shows that the
space of the physical motions is very limited: it is 
 just a
two-dimensional subspace of the infinite dimensional space of all motions.  
Only a two-dimensional space of curves $\gamma$ is realized in nature.

In the case of the small oscillations of a frictionless pendulum, we can
coordinatize the physical motions by the two real numbers $A\ge 0$ and
$0\le\phi<2\pi$, and (\ref{uno}) is given by
\begin{equation}
    f(\alpha,\beta;A,\phi)=\alpha - A\sin(\omega \beta+\phi) =0.
    \label{eq:f2}
\end{equation}
This equation gives a curve $\gamma$ in $\cal C$ for each couple $(A,\phi)$.
Equation (\ref{eq:f2}) is the mathematical law that captures the entire
empirical information we have on the dynamics of the pendulum.  

Let $\Gamma$ be the two-dimensional space of the physical motions,
coordinatized by $A$ and $\phi$.  $\Gamma$ is the {\em relativistic
phase space\/}\index{phase space!relativistic} of the pendulum (or the
{\em space of the motions\/}).  A point in $\Gamma$, is a
{\em relativistic state}. 

A relativistic state is determined by a couple $(A,\phi)$.  It
determines a curve $\gamma$ in the $(\alpha,\beta)$ plane.  That is, it
determines a correlations between the two partial observables $\alpha$
and $\beta$, via (\ref{eq:f2}). 

\subsection{General structure of the dynamical
systems.}\label{genstr}

The $({\cal C}, \Gamma, f)$ language described above is general.  On
the one hand, it is sufficient to describe all predictions of
conventional mechanics.  On the other hand, it is broad enough to
describe general relativistic systems.    All fundamental systems, including
the general relativistic ones, can be
described (to the accuracy at which quantum effects can be
disregarded) by making use of these concepts:
\begin{enumerate}
     \item[(i)] The relativistic \emph{configuration space} $\cal C$,
     of the partial observables.  
     \item[(ii)] The relativistic \emph{phase
     space} $\Gamma$ of the
     relativistic \emph{states}.  \item[(iii)] The \emph{evolution
     equation} $f=0$, where $ f: {\Gamma}\times{\cal C} \to V$.
 \end{enumerate}
$V$ is a linear space.  The state in the phase space $\Gamma$ is fixed until the
system is disturbed.  Each state in $\Gamma$ determines (via $f=0$) a motion
$\gamma$ of the system, namely a relation, or a set of relations, between the
observables in $\cal C$.\footnote{A motion is not necessarily a one-dimensional 
curve in $\cal C$: it can be a surface in $\cal C$ of any dimension $k$.  
If $k>1$, we say that there is gauge invariance. Here I take $k=1$.  See \cite{book} for 
the general case with gauge invariance.}
Once the state is determined (or guessed), the
evolution equation predicts all the possible events, namely all the allowed
correlations between the observables, in any subsequent measurement.

Notice that this language makes no reference to a special ``time" variable.
The definitions of observable, state, configuration space and phase space given
here are different from the conventional definition.  In particular, notions of
instantaneous state, evolution in time, observable at a fixed time, play no role
here.  These notions make no sense in a general relativistic context.  

\subsection{Hamiltonian mechanics}

It appears that all elementary physical systems can be described by hamiltonian
mechanics.\footnote{Perhaps because they are the classical limit of a
quantum system.} Once the kinematics --that is, the space $\cal C$ of
the partial observables $q^a$-- is known, the dynamics --that is,
$\Gamma$ and $f$-- is fully determined by giving a surface $\Sigma$ in
the space $\Omega$ of the observables $q^a$ and their momenta $p_{a}$. 
The surface $\Sigma$ can be specified by giving a function $H: \Omega
\to R$.  \index{hamiltonian!relativistic} $\Sigma$ is then define
 d
by $H=0$.\footnote{Different $H$'s that vanish on the same surface
$\Sigma$ define the same physical system.} Denote $\tilde\gamma$ a
curve in $\Omega$ (observables and momenta) and $\gamma$ its
restriction to $\cal C$ (observables alone).  $H$ determines the
physical motions via the following

\begin{quotation}
\noindent {\em {\bf \em Variational principle.} A curve $\gamma$ connecting the
events $q_{1}^a$ and $q_{2}^a$ is a physical motion if $\tilde\gamma$ extremizes
the action \index{variational principle} 
$
     S[\tilde\gamma] = \int_{\tilde\gamma}\ p_{a}\; dq^a
	      \label{S}
     $
in the class of the curves $\tilde\gamma$ satisfying
     $
	  H(q^a,p_{a})=0
	  \label{H0}
	$
whose restriction $\gamma$ to $\cal C$ connects $q_{1}^a$ and $q_{2}^a$.}
\end{quotation}
All known physical (relativistic and nonrelativistic) hamiltonian systems can be formulated in
this manner.  Notice that no notion of time has been used in this formulation. 
I call $H$ the \emph{relativistic hamiltonian}, or, if
there is no ambiguity, simply the \emph{hamiltonian}.  I denote the pair $({\cal
C},H)$ as a {\em relativistic dynamical system}.  

A nonrelativistic system is a system where one of the partial observables,
called $t$,  is singled out as playing a special role. And the Hamiltonian has
the particular structure 
\begin{equation}
H=p_t+H_0(q_i,p^i,t)
\label{nonr}
\end{equation}
where $p_t$ is the momentum conjugate to 
$t$ and $(q_i,p^i)$ are the other variables. 
This structure is {\em not} necessary in order to have a
well-defined physical interpretation of the formalism.  The relativistic hamiltonian $H$ is related to, but should not be
confused with, the usual nonrelativistic hamiltonian $H_{0}$. In the case of the pendulum,
for instance, $H$ has the form
\begin{eqnarray} 
H=p_\beta+\frac{1}{2m}\ p_\alpha^2
+\frac{m\omega^2}{2}\ \alpha^2. 
\label{Hpendulum2}
\end{eqnarray} 
 $H$ always exists, while $H_{0}$ exists only for the
non-general-relativistic systems.

\section{Formal structure of timeless quantum mechanics}

A formulation of QM slightly more general than the conventional one --
 or a quantum version of the relativistic classical mechanics
 discussed above-- is needed to describe systems where no preferred time variable is specified.  Here I sketch the possibility of such a formulation.   For a detailed discussion of the issues raised by this formulation, see \cite{book}.  The quantum theory can be defined in terms of the following quantities.
 
\begin{description} 
\addtolength{\itemsep}{-.1cm}

\item[\rm\em Kinematical states.] 
Kinematical states form a space $\cal S$ in a rigged Hilbert space
${\cal S}\subset{\cal K}\subset{\cal S}'$.

\item[\rm\em Partial observables.]  A partial observable is
represented by a self-adjoint operator in $\cal K$.  Common
eigenstates $|s\rangle$ of a complete set of commuting partial
observables are denoted quantum events.\index{observable!partial}

\item[\rm\em Dynamics.]  Dynamics is defined by a self-adjoint
operator $H$ in $\cal K$, the (relativistic) hamiltonian. 
\index{hamiltonian!relativistic} The operator from $\cal S$ to ${\cal
S}'$
\begin{equation} 
P = \int d\tau \ e^{-i\tau H} 
\label{Puno}
\end{equation} 
is (sometimes improperly) called the projector.  \index{projector}
(The integration range in this integral depends on the system.)  Its
matrix elements
\begin{equation} 
W(s,s') = \langle s |P|s'\rangle 
\label{Pss}
\end{equation} 
are called transition amplitudes.
  \index{transition amplitudes}

\item[\rm\em Probability.]  Assume discrete spectrum for simplicity. The probability of the
quantum event $s$ given the quantum event $s'$ is 
\begin{equation}
{\cal P}_{ss'} = |W(s,s')|^2 
\end{equation} where $|s\rangle$ is
normalized by $\langle s |P|s\rangle=1$. 
 \end{description} 

To this we may add: 
 \begin{description} 
\addtolength{\itemsep}{-.1cm}
 \item[\rm\em States.]  A physical state is a solution of the
Wheeler-DeWitt equation 
\begin{equation}
H\psi=0.
\label{WdWeq}
\end{equation}
Equivalently, it is an element of the Hilbert space $\cal H$ defined
by the quadratic form $\langle \ \cdot\ |P| \ \cdot\ \rangle$ on $\cal
S$.  

\item[\rm\em Complete observables.]  A complete observable $\cal A$ is
represented by a self-adjoint operator on $\cal H$.  A self-adjoint
operator $A$ in $\cal K$ defines a complete observable if
\index{observable!complete}
\begin{equation}
[A,H]=0. 
\end{equation}

\item[\rm\em Projection.]  If the observable $\cal A$ takes value in
the spectral interval $I$, the state $\psi$ becomes then the state
$P_{I}\psi$, where $P_I$ is the spectral projector on the interval
$I$.  If an event corresponding to a sufficiently small region $R$ is
detected, the state becomes $|R\rangle$.
\end{description} 

A relativistic quantum system is defined by a rigged Hilbert space of
kinematical states $\cal K$ a
 nd a set of partial observables $A_{i}$
including a relativistic hamiltonian operator $H$.  Alternatively, it
is defined by giving the projector $P$.
 
The
structure defined above is still tentative and perhaps incomplete. 
There are aspects of this structure that deserve to be better
understood, clarified, and specified.  Among these is the correct 
treatment of repeated
measurements.  On the other hand, the conventional structure of QM is
\emph{certainly} physically incomplete, in the light of GR. The above
is an attempt to complete it, making it general relativistic.

\subsection{Quantization and classical limit}

If we are given a classical system defined by a non relativistic 
configuration space ${\cal C}$ with coordinates $q^a$ and by a 
relativistic hamiltonian $H(q^a,p_{a})$, a solution of the quantization 
problem is provided by the multiplicative operators $q^a$, the derivative 
operators 
$ p_{a}=-i\hbar\frac{\partial}{\partial q^a}
$
and the 
Hamiltonian operator 
$       H= H\left(q^a,-i\hbar\frac{\partial}{\partial q^a}\right) 
$ on the Hilbert space ${\cal K}=L^2[{\cal C},dq^a]$, or more precisely,
the Gelfand triple determined by $\cal C$ and the measure $dq^a$.  The
physics is entirely contained in the transition amplitudes
\begin{equation} 
        W(q^a,q'{}^a)=\langle q^a| P|q'{}^a\rangle 
\end{equation} 
where the 
states $|q^a\rangle$ are the eigenstates of the multiplicative operators 
$q^a$. 
In the limit $\hbar\to 0$ the Wheeler-DeWitt equation becomes the
relativistic Hamilton-Jacobi equation (\cite{book}). 
 
An example of relativistic formalism is provided
by the quantization of the pendulum described in the previous section:
The kinematical state space is ${\cal K}=L_{2}[R^2,d\alpha\; d\beta]$.  The
partial observable operators are the multiplicative operators $\alpha$
and $\beta$ acting on the functions $\psi(\alpha,\beta)$ in ${\cal K}$. 
Dynamics is defined by the operator $H$ given in (\ref{Hpendulum2}). 
The Wheeler-DeWitt equation becomes the Schr\"odinger equation. 
$\cal H$ is the space of solutions of this equation.  The ``projector"
operator $P: {\cal K}\to {\cal H}$ defined by $H$ defines the scalar 
product in $\cal H$.  Its matrix
elements $W(\alpha,t,\alpha',t')$ between the common eigenstates of
$\alpha$ and $t$ are given by the oscillator's propagator.  They
express all predictions of the theory.  Because of the specific form
of $H$, these define a probability density in $\alpha$ but not in $\beta$.\footnote{
For more details, see \cite{book}. For a related approach, see 
\cite{hartle}, \cite{Halliwell}.} This example is of course trivial, since the system admits
also a conventional hamiltonian quantization.   But the point here is that the
formalism above remains viable also for general relativistic systems that do 
not admit a conventional hamiltonian quantization, because they do not have 
a preferred time variable.  

\section{Recovery of time}

If we formulate the fundamental theory of nature in a timeless language, we have then
the problem of recovering the familiar notion of time.  There are two aspects of this
problem.  
The first aspect is simply to see if the general theory admits a regime or an approximation where the Hamiltonian can be approximated by an Hamiltonian of the form (\ref{nonr}) for some partial observable $t$.   If this is the case, the system is described within this regime and this approximation precisely as a standard nonrelativistic system, and we can therefore identify $t$ with the nonrelativistic time. 

The second aspect is far more subtle.   The time of our experience is associated with a number of peculiar features that make it a very special physical variable.   Intuitively (and imprecisely) speaking, time ``flows", we can never ``go back in time", we remember the past but not the future, and so on.  Where do all these very peculiar features of the time variable come from?

I think that these features are not mechanical. Rather they emerge at the thermodynamical level.  More precisely, these are all features that emerge when we give an approximate statistical description of a system with a large number of degrees of freedom. 
We represent our incomplete knowledge and assumptions in terms of a
statistical state $\rho$.  The state $\rho$
can be represented as a normalized positive function on the relativistic phase
space $\Gamma$
\begin{eqnarray}
&&\ \rho:\ \ \ \Gamma \ \ \to\ \ R^+,\\ 
&&\int_{\Gamma} ds \ \rho(s)\ \  =\ \ 1.
\label{thermalstate}
\end{eqnarray}
$\rho(s)$ represents the assumed probability density of the state $s$
in $\Gamma$.  Then the expectation value of any observable
$A:\Gamma\to R$ in the state $\rho$ is
\begin{eqnarray}
\rho[A] & = & \int_{\Gamma} ds \ A(s)\ \rho(s). 
\label{thermalaverage}
\end{eqnarray} 
The fundamental postulate of statistical mechanics is that a system
left free to thermalize reaches a time independent equilibrium state
\index{state!equilibrium} that can be represented by means of the
Gibbs statistical state
\begin{eqnarray}
\rho_{0}(s) & = & N \ e^{-\beta H_{0}(s)}.
\label{gibbs}
\end{eqnarray}
where $\beta=1/T$ is a constant --the inverse temperature-- 
 and
$H_{0}$ is the nonrelativistic hamiltonian.  Classical thermodynamics
follows from this postulate.   Time evolution  is determined by 
$A_{t}(s) = A(t(s))$ where $s(t)$ is the hamiltonian flow of $H_{0}$
on $\Gamma$.  The correlation probability between $A_{t}$ and $B$ is
\begin{eqnarray}
W_{AB}(t)& = & \rho_{0}[\alpha_{t}(A)B]\ = \ \int_{\Sigma} ds \
A(s(t))\ B(s)\ e^{-\beta H_{0}(s)}.
\label{AB} 
\end{eqnarray}

I have argued that mechanics does not
single out a preferred variable, because all mechanical predictions
can be obtained using the relativistic hamiltonian $H$, which treats
all variables on equal footing.  Is this true also for
statistical mechanics?  Eqs
(\ref{thermalstate}-\ref{thermalaverage}) are meaningful also in the
relativistic context, where $\Gamma$ is the space of the solutions of
the equations of motion.  But this is not true for 
(\ref{gibbs}) and (\ref{AB}).  These depend on the nonrelativistic
hamiltonian.  They single out $t$ as a special variable.  
With purely mechanical measurement we cannot recognize the time
variable.  With statistical or thermal measurements, we can.
In principle, we can figure out $H_{0}$ simply
by repeated microscopic measurements on copies of the system, without
any need of observing time evolution.  Indeed, if we find out the
distribution of microstates $\rho_{0}$, then, up to an irrelevant
additive constant we have
\begin{eqnarray}
H_0 = -\frac{1}{\beta}\ \ln \rho_{0}. 
\label{gibbsinverso}
\end{eqnarray}
Therefore, in a statistical context we have in principle an
operational procedure for determining which one is the time variable: 
Measure $\rho_{0}$; compute $H_{0}$ from
(\ref{gibbsinverso}); compute the hamiltonian flow $s(t)$ of
$H_{0}$ on $\Sigma$: the time variable $t$ is the parameter of this
flow.   The multiplicative constant in front of
$H_{0}$ just sets the unit in which time is measured.  Up to this
unit, we can find out which one is the time variable just by measuring
$\rho_{0}$. 
  This is in contrast with the purely mechanical
context, where no operational procedure for singling out the time
variable is available. 

Now, consider a truly
relativistic system where no partial observable is singled out as the
time variable.  We find that the statistical state describing the system
is given by a certain \emph{arbitrary}\footnote{For the
(\ref{gibbsinverso2}) to make sense, assume that $\rho$ nowhere
vanishes on $\Sigma$.\label{footn}} state $\rho$.  \emph{Define} the quantity
\begin{eqnarray}
H_{\rho} = - \ln \rho.
\label{gibbsinverso2}
\end{eqnarray}
Let $s(t_{\rho})$ be the hamiltonian flow of $H_{\rho}$.  Call
$t_{\rho}$ ``thermal time".  \index{time!thermal} Call ``thermal
clock" any measuring devise whose reading grows linearly with this
flow.  Given an observable $A$, consider the one-parameter family of
observables $A_{t_{\rho}}$ defined by 
$A_{t_{\rho}}(s)=A(t_{\rho}(s))$.  Then it follows that the
correlation probability between the observables
  $A_{t_{\rho}}$ and $B$
is
\begin{eqnarray}
W_{AB}(t_{\rho})& = & \int_{\Sigma} ds \ A(t_{\rho}(s)) B(s)
e^{-H_{\rho}(s)}.
\label{ABt} 
\end{eqnarray}
Notice that there is no  difference between the physics described by 
(\ref{gibbs}-\ref{AB}) and the one described by
(\ref{gibbsinverso2}-\ref{ABt}).  \emph{Whatever}
the statistical state $\rho$ is, there exists always a variable
$t_{\rho}$, measured by the thermal clock, \index{thermal clock} with
respect to which the system is in equilibrium and physics is the same
as in the conventional nonrelativistic statistical case !  

This observation leads us to the following hypothesis.
\begin{quote}
{\bf\em The thermal time hypothesis.} 
In nature, there is no preferred physical time variable $t$.  There
are no equilibrium states $\rho_{0}$ preferred a priori.  Rather, all
variables are equivalent; we can find the system in an arbitrary state
$\rho$; if the system is in a state $\rho$, then a preferred variable
is singled out by the state of the system.  This variable is what we
call time.
\end{quote}
In other words, it is the statistical state that determines which
variable is physical time, and not any a priori hypothetical ``flow"
that drives the system to a preferred statistical state. 
When we say that a certain variable is ``the time", we
are not making a statement concerning the fundamental mechanical
structure of reality.  Rather, we are making a statement about the
statistical distribution we use to describe the macroscopic properties
of the system that we describe macroscopically.
The ``thermal time hypothesis" is the idea that
what we call ``time" is the thermal time of the statistical
state in which the world happens to be, when described in terms of the
macroscopic parameters we have chosen.  

Time is, that is to say, the expression of our ignorance of the full 
microstate.\footnote{It seems to me that the general point of view
I am adopting here might be compatible with the perspectival one
presented by Jenann Ismael in \cite{J} and, recently, in 
\cite{J2}.}

The thermal time hypothesis works surprisingly well in a number of
cases.  For example, if we start from radiation filled covariant
cosmological model, with no preferred time variable and write a
statistical state representing the cosmological background radiation,
then the thermal time of this state turns out to be precisely the
Friedmann time \cite{RovelliThermalTime2}.  Furthermore, 
this hypothesis extends in a very 
natural way to the quantum context, and even more naturally to the
quantum field theoretical context, where it leads also to a general
abstract state-independent notion of time flow.
In QM, the time flow is
given by
\begin{eqnarray}
A_{t}= \alpha_{t}(A) = e^{it H_{0}}A e^{-it H_{0}} .
\label{timeflowQ} 
\end{eqnarray} 
A statistical state \index{state!statistical} is described by a
density matrix $\rho$.  It determines the expectation values of any
observable $A$ via
\begin{eqnarray}
\rho[A] =  Tr[A\rho].
\end{eqnarray} 
This equation defines a positive functional $\rho$ on the observables'
algebra. The relation between a quantum
Gibbs state $\rho_{0}$ and $H_{0}$ is the same as in equation
(\ref{gibbs}).  That is
\begin{eqnarray}
\rho_{0}& = & N\ e^{-\beta H_{0}}. 
\label{rhoHzeroQ}
\end{eqnarray}
Correlation probabilities can be written as
\begin{eqnarray}
W_{AB}(t)& = & \rho[\alpha_{t}(A)B] = Tr[e^{it H_{0}}A e^{-it H_{0}} B
e^{-\beta H_{0}}], 
\label{ABtQ}
\end{eqnarray}
Notice that it follows immediately from the definition that  
\begin{eqnarray} 
\rho_{0}[\alpha_{t}(A)B]& = & \rho_{0}[\alpha_{(-t-i\beta)}(B)A], 
\label{KMS}
\end{eqnarray}
Namely 
\begin{eqnarray} 
W_{AB}(t) = W_{BA}(-t-i\beta)
\label{KMS2}
\end{eqnarray}
A state $\rho_{0}$ over an algebra, satisfying the relation
(\ref{KMS}) is said to be $KMS$ with respect to the flow $\alpha_{t}$.

We can now generalize the thermal time hypothesis.  Given a generic
state $\rho$ the thermal hamiltonian is defined by
\index{time!thermal}
\begin{eqnarray}
H_{\rho} = -\ln \rho
\label{Htr}
\end{eqnarray}
and the thermal time flow is defined by 
\begin{eqnarray}
A_{t_{\rho}}& = & \alpha_{t_{\rho}}(A) \ =\ e^{it_{\rho} H_{\rho}}A
e^{-it_{\rho} H_{\rho}}.
\label{ABtQM} 
\end{eqnarray}
$\rho$ is a KMS state with respect to the thermal time flow  it defines.

In QFT, finite
temperature states do not live in the same Hilbert space as the zero
temperature states.  $H_{0}$ is a divergent operator on these states. 
Therefore equation (\ref{rhoHzeroQ}) make no sense.
But Gibbs states can still be characterized by (\ref{KMS}): a Gibbs state
$\rho_{0}$ over an algebra of observables is a KMS state with respect
to the time flow $\alpha(t)$.

A celebrated theorem by Tomita states precisely that
given any\footnote{Any \emph{separating} state $\rho$.  A separating
density matrix has no zero eigenvalues.  This is the QFT equivalent of
the condition stated in the footnote 9.} state $\rho$ over a von Neumann 
algebra\footnote{The
observable algebra is in general a $C^*$ algebra.  We obtain a von
Neumann algebra by closing in the Hilbert norm of the quantum state
space.}, there is always a flow $\alpha_{t}$, called the Tomita flow
of $\rho$, such that (\ref{KMS}) holds.
This theorem allows us to extend (\ref{gibbsinverso2}) to QFT: the
thermal time flow $\alpha_{t_{\rho}}$ is defined in general as the
Tomita flow of the statistical state $\rho$ \cite{ConnesRovelli,Diamonds}.
Thus the thermal times hypothesis can be readily extended to QFT. What
we call the flow of ``time" is the Tomita flow of the
statistical state $\rho$ in which the world happens to be, when
described in terms of the macroscopic parameters we have 
chosen.\footnote{A von Neumann algebra posses also a more
abstract notion of time flow, independent from $\rho$.  This is given
by the one-parameter group of outer automorphisms, formed by the
equivalence classes of of automorphisms under inner (unitary)
automorphisms.  Alain Connes has shown that this group
is independent from $\rho$.  It only depends on the algebra itself. 
Connes has stressed the fact that this group provides an abstract
notion of time flow that depends only on the algebraic structure of
the observables, and nothing else.}

\section{Conclusion}

I have presented a certain number of ideas and results:
\begin{enumerate}

\item It is possible to formulate classical mechanics in a way in which the time
variable is treated on equal footings with the other physical variables, and not singled
out as the special independent variable.  I have argued that this
is the natural formalism for describing general relativistic systems. 

\item It is possible to formulate quantum mechanics in the same manner. I think that
this may be the effective formalism for quantum gravity. 

\item The peculiar properties of the time variable are of thermodynamical origin, and
can be captured by the thermal time hypothesis. Within quantum field theory, 
``time" is  the Tomita flow of the statistical state $\rho$ in which the world happens 
to be, when described in terms of the macroscopic parameters we have chosen

\item In order to build a quantum theory of gravity the most effective strategy is therefore
to forget the notion of time all together, and to define a quantum theory capable of 
predicting the possible correlations between partial observables. 

\end{enumerate}

Before concluding, I must add that the views expressed are far from being entirely 
original.  I have largely drawn from the ideas of numerous scientists, and in particular
Bryce DeWitt, John Wheeler, Chris Isham, Abhay Ashtekar, Jorge Pullin, Rodolfo 
Gambini,  Don Marolf, Don Page, Bianca Dittrich, Julian Barbour and Karel Kuchar, 
William Wootters, Jean-Marie Souriau, Lee Smolin, John Baez, Jonathan Halliwell, 
Jim Hartle, Alain Connes, and certainly others that I forget  here.  I have here attempted 
to combine a coherent view about the problem of time in quantum gravity, starting from 
what others have understood. 

On the other hand, I also see well that the view I present here is far from being uncontroversial.
Several authors maintain the idea that the notion of time is irreducible, and cannot be eliminated from fundamental physics. See for instance \cite{Smolin:2000vs}. I could of course be wrong, but my own expectation is that the notion of time is extremely natural to us, but only in the same manner in which other intuitive ideas are rooted in our intuition because they are features of the small garden in which we are accustomed to living (for instance: absolute simultaneity, absolute velocity, or the idea of a flat Earth and an absolute up and down).  Intuition is not a good guide for understanding natural regimes so distant from our daily experience. The best guide is provided by the theories of the world that have proven empirically effective, and therefore summarize the knowledge we have about Nature.  In particular, general relativity challenges strongly our intuitive notion of a universal flow of time. I think we must take its lesson seriously.

\appendix

\section{Geometric formalism.}\label{presymplectic}

The timeless mechanical formalism can be expressed in a beautiful and compact manner
using the geometric form.  The variables $(q^a, p_{a})$ are
coordinates on the cotangent space $\Omega=T^*{\cal C}$. Equation $H=0$ 
defines a surface $\Sigma$ in this space. The cotangent space
carries the natural one-form
\begin{equation}
\tilde\theta=p_{a}dq^a.
\end{equation}
Denote $\theta$ the restriction of $\tilde\theta$ to the surface $\Sigma$.  The
two-form $\omega=d\theta$ on $\Sigma$ is degenerate: it has null directions. 
The integral surfaces of these null directions are the \emph{orbits} of $\omega$
on $\Sigma$.  Each such orbit projects from $T^*{\cal C}$ to $\cal C$ to give a
surface in $\cal C$.  These surfaces are the motions.

$\Sigma$ has dimension $2n-1$, the kernel of $\omega$ is, generically, one-dimensional, and the motions are, generically, one-dimensional.  Let $\tilde\gamma$, be a motion on $\Sigma$
and $X$ a vector tangent to the motion.  Then
\begin{equation}
	   \omega(X)=0.
    \label{eq:omegaX}
\end{equation}
To find the motions, we have just t
 o integrate this equation.  Equation
(\ref{eq:omegaX}) is the equation of motion.  $X$ is defined by the homogeneous
equation (\ref{eq:omegaX}) only up to a multiplicative factor.  Therefore the
tangent of the orbit is defined only up to a multiplicative factor, and
therefore the parametrization of the the orbit is not determined by equation
(\ref{eq:omegaX}): what matters is only the \emph{relation} that the orbits establish
between the different variables.

\end{document}